# Nuclear Shape Transition, Triaxiality and Energy Staggering of ϒ-Band States for Even-Even Xenon Isotopic Chain

W. B. Elsharkawy[1,*], Abeer Mera[2,3], M. Kotb[4] and A.M. Khalaf[4]

[1]Physics Department, College of Science and Humanities, Prince Sattam bin Abdulaziz University, 11942 Alkharj, Saudi Arabia, KSA.
[2]Physics Department, College of Arts and Science, , Prince Sattam Bin Abdulaziz University, Wadi Addawasir, 11991, Kingdom of Saudi Arabia.
[3]Physics Department, Faculty of Science, Kafrelsheikh University, Kafrelsheikh 33516; Egypt.
[4]Physics Department, Faculty of Science, Al-Azhar University, Cairo, Egypt
*Corresponding author: Drwafaa8@gmail.com

## Abstract

The positive-parity states of even-even Xe nuclei are investigated within the framework of modified O(6) limit of the interacting boson model (IBM1). The effective three-body interaction [QQQ]$^{(0)}$ where Q is the IBM O(6) quadrupole operator is introduced to exhibit the triaxiality nature. The shape of nuclear surface is described by the deformation parameters β,ϒ by using the intrinsic coherent state. The potential energy surfaces (PES,s) of the transition U(5)-Triaxiality-O(6) are calculated and analyzed and the critical phase transition points are identified. For each nucleus a fitting procedure is adopted to get the best model parameters by fitting some selected calculated energy levels and B(E2) transition rates ratios with experimental ones. These ratios are analyzed because they serve as effective order parameters in the shape phase transition. The nuclei in Xe isotopic chain evolve from spherical vibrator U(5) to ϒ-soft rotor O(6) by increasing the boson number from N=3 (heavy isotope $^{132}$Xe) to N=10 (light isotope $^{120}$Xe) and the isotope $^{126}$Xe represent the critical nucleus. The nucleus $^{128}$Xe has triaxial nature. To deal with high spin states in ϒ-band in $^{118-128}$Xe isotopic chain to investigate and exhibit the odd-even spin energy staggering, we introduce the two parameters collective nuclear softness rotor model (CNS2). Three different staggering indices depending on the dipole transitions linking the two families of spins and the quadrupole transitions within each spin family are considered. Strong odd-even spin energy staggering has been seen. As a link between the IBM and CNS2 models we observed that the energy difference $[E(I_\Upsilon^+) - E(I_g^+)]$ between the ϒ-band and ground state band normalized to $[E(2_\Upsilon^+) - E(2_g^+)]$ decreases with increasing the mass number.

## 1. Introduction

Structures, electromagnetic transition rates and shape phase transitions (which describe how one geometric shape evolve into another) of several collective nuclei in different mass regions were systematically studied and analyzed in framework of the algebraic interacting boson model (IBM1)[1-12]. In IBM, collective excitations in nuclei are described in terms of a system of N interacting monopole (s) and quadrupole ($d_m$ with m=0, $\pm$1, $\pm$2) bosons with angular momentum L=0 and L=2 respectively. In its simplest version the model does not distinguish between protons and neutrons and have U(6) group. There are three possible phases known as U(5),Su(3) and O(6) corresponding to the three limiting symmetries of the geometric collective model(GCM) [13-17] namely: spherical vibrator, axially symmetric deformed rotor and $\gamma$-unstable respectively.

The critical point symmetries E(5),X(5),Y(5) and Z(5) were introduced [18-21] to describe the critical point of the phase transition from spherical vibrator to $\gamma$-unstable rotor, from spherical vibrator to axial symmetric rotor, from axially deformed shapes to triaxail deformed shapes and for the prolate to oblate nuclear shape transition respectively. The study of collective Bohr-Hamiltonian near the critical point symmetries in nuclei is an important topic of nuclear structure [22-25]. Also the critical point symmetries are needed to study the transition between two different shapes [26-28].

The connection between the IBM, potential energy surfaces (PES,s),geometric shapes and phase transitions can be obtained by introducing the intrinsic coherent state[29]. The shape of deformed nuclei can be classified into prolate, oblate and triaxial nuclei [30] according to the ratios between the three principle axes of rotation in the ellipsoid. The IBM1 in its sd-version and with only up to two-body interactions cannot create stable triaxial shapes [31, 32], also IBM1 cannot identify the high-lying states in $\gamma$-bands. A three-body interaction terms were added to the IBM1 Hamiltonian to produce the effect of triaxialty on the energy spectrum [33-40].

The triaxial shapes together with softness with respect to asymmetry parameter γ in the light rare-earth region in A~130 mass region was investigated, so that the present study deals also with softness. By treating the variation of the moment of inertia with the nuclear spin, the nuclear softness parameters are defined [41]. Keeping the nuclear softness parameters to only first order, the nuclear softness model is denoted by soft rotor formula (SRF) [42] or collective nuclear softness (CNS2) model, because the model contains only two parameters: the ground state moment of inertia and the nuclear softness parameter. The NS model was remarkable successful in the description of the ground and γ bands of many medium and heavy even-even nuclei. The triaxilty and softness in Xe, Ba and Te isotopic chains were studied experimentally [43-47] and interpreted by different nuclear models [48-52].

In The present paper the effects of adding the three [QQQ]$^{(0)}$ interaction (where Q is the O(6) symmetric quadrupole operator) to the dynamical symmetry O(6) of the IBM1 Hamiltonian is studied to generate rotational and triaxial nuclear states without the use of the rotational Su(3) limit to IBM1. Analysis of PES,s will give the evolution of shape transition in Xenon isotopic chain. Furthermore, we present here also the CNS2 model to investigate the ΔI=1 odd-even spin staggering in γ-bands of the even-even $^{120-134}$Xe isotopic chain. There is poor information about the ϒ-bands of very light and very heavy Xe nuclei. This pays attention to estimate the available measured ϒ-band energy levels of the intermediate isotopes $^{120-126}$Xe by CNS2 model, while the extended IBM is used for the calculations of ground state bands.

## 2. Outline of the IBM Approach

We choose the IBM Hamiltonian operator of the O(6) dynamical symmetry arises from the group chain U(6) ⊃O(6) ⊃O(5) ⊃ O(3) in terms of multipole operators as:

$$H[O(6)] = a_0 \hat{P}^\dagger . \hat{P} + a_1 \hat{L} . \hat{L} + a_3 \hat{T}_3 . \hat{T}_3 \qquad (1)$$

where $\hat{P}, \hat{L}, \hat{T}_3$ are the pairing operator, the angular momentum operator and the octupole operator defined by:

$$\hat{P} = \frac{1}{2}(\tilde{d}.\tilde{d}) - \frac{1}{2}(\tilde{s}.\tilde{s}) \tag{2}$$

$$\hat{L} = \sqrt{10}\,[d^\dagger \times \tilde{d}]^{(1)} \tag{3}$$

$$\hat{T}_3 = [d^\dagger \times \tilde{d}]^{(3)} \tag{4}$$

with $(s^\dagger, d^\dagger)$ and $(\tilde{s}, \tilde{d})$ are the creation and annihilation operators. The symbol dot (.) stands for the scalar product defined as

$$\hat{T}_L . \hat{T}_L = \sum_M (-1)^M \hat{T}_{L,M} . \hat{T}_{L,-M} \tag{5}$$

where $\hat{T}_{L,M}$ corresponds to the component of the operator $\hat{T}_L$. The operator $\hat{d}_m = (-1)^m \hat{d}_{-m}$ and $\tilde{s} = s$ are introduced to ensure the correct tensorial character under spatial rotations.

The corresponding energy eigenvalues of the Hamiltonian (1) are

$$E_{O(6)}(N, \sigma, \tau, I) = \frac{1}{4}A(N-\sigma)(N+\sigma+4) + B\tau(\tau+3) + CI(I+1) \tag{6}$$

with A = $a_0$, B = $a_3$, C = $a_1$-(1/10) $a_3$

where [N], σ, τ, I are the quantum numbers which distinguish U(6), O(6), O(5) and O(3) respectively and for ground state band N=σ.

The IBM being an algebraic model, it does not contain shape variables, so that the geometrical interpretation of the IBM Hamiltonian can be obtained by introducing the intrinsic coherent state formalism [29] in terms of shape deformation parameters β, γ by considering the state:

$$|N\beta\gamma\rangle = \frac{1}{\sqrt{N!}}(b_c^\dagger)^N|0\rangle \tag{7}$$

where $|0\rangle$ is the boson vacuum and $b_c^\dagger$ is the boson creation operator given by

$$b_c^\dagger = \frac{1}{\sqrt{1+\beta^2}}\left[s^\dagger + \beta\cos\gamma\, d_0^\dagger + \frac{1}{\sqrt{2}}\beta\sin\gamma\,(d_2^\dagger + d_{-2}^\dagger)\right] \tag{8}$$

The potential energy surface (PES) is defined as the expectation value of the IBM Hamiltonian

$$E(N\beta\gamma) = \frac{\langle N\beta\gamma|H|N\beta\gamma\rangle}{\langle N\beta\gamma|N\beta\gamma\rangle} \tag{9}$$

Using the Hamiltonian equation (1), the PES becomes

$$E[O(6)] = \frac{A_2\beta^2 + A_4\beta^4}{(1+\beta^2)^2} + A_0 \tag{10}$$

where $A_2 = [\lambda - a_0(N-1)]N$, $A_4 = \lambda N$, $A_0 = \frac{1}{4}a_0(N-1)]N$, $\lambda = 6a_1 + \frac{7}{5}a_3$

The PES equation (10) is γ-independent. The equilibrium value of β is given by

$$\beta_e = \pm\sqrt{\frac{a_0(N-1)-\lambda}{a_0(N-1)+\lambda}} \text{ for } \lambda < a_0(N-1) \tag{11}$$

The antispinodal point is found at λ=$a_0$(N-1) and the PES at this point is given by

$$E_{critical} = \frac{\lambda N \beta^4}{(1+\beta^2)^2} + \frac{1}{4}a_0 N(N-1) \tag{12}$$

For pure O(6) limit we take only the pairing term (i.e. λ=0) and the equilibrium deformation parameter β becomes $\beta_e$=1.

In Figure (1) we illustrated the behavior of the PES corresponding to the O(6) Hamiltonian equation (1) for fixed $a_0$ at 101.2 and varying λ to produce critical point at N=7. It is clear that the N effect is important.

To remove the dependence on the boson number N for classical limit, we rewrite Hamiltonian (1) in the form

$$\hat{H} = \frac{a_0}{N^2}\hat{P}^\dagger \cdot \hat{P} + \frac{a_1}{N}\hat{L}\cdot\hat{L} + \frac{a_3}{N}\hat{T}_3\cdot\hat{T}_3 \tag{13}$$

The corresponding PES reads

$$E(\beta) = \frac{(\lambda-a_0)\beta^2 + \lambda\beta^4}{(1+\beta^2)^2} + \frac{1}{4}a_0 \tag{14}$$

One can see that for $a_0 \leq 0$, the deformation parameter β is at β=0, for $a_0 > \lambda$ we reach the deformed γ-soft shape, which for $a_0 = \lambda$ a shape transition from spherical to γ-unstable occurs. Figure (2) illustrate the behavior of the PES's at λ=4 and $a_0$-values $a_0$=0,2,4,8,12. A shape transition occurs at $a_0$= λ=4.

### 3. Extended O(6) IBM

The aim of the present paper is to exhibit the shape phase transition and nuclear triaxiality from O(6) dynamical symmetry of the IBM. The transition between the spherical U(5) and γ-unstable O(6) shape can be studied by adding the term $\varepsilon \hat{n}_d$ to the Hamiltonian (1) where $\hat{n}_d$ is the d-boson number operator defined as $\hat{n}_d = d^\dagger \cdot \tilde{d}$

$$H = H[O(6)] + \varepsilon \hat{n}_d \tag{15}$$

By using the intrinsic coherent state equation (8), the corresponding PES as a function of deformation parameter β is given by

$$E(N,\beta) = \frac{A_2^\backslash \beta^2 + A_4^\backslash \beta^4}{(1+\beta^2)^2} + A_0 \quad where \quad A_2^\backslash = [\lambda + \varepsilon - a_0(N-1)]N, \, A_4^\backslash = (\lambda + \varepsilon)N \tag{16}$$

In order to exhibit the triaxiality, we will add to the O(6) IBM Hamiltonian the cubic interaction $[QQQ]^{(0)}$ where Q is the O(6) quadrupole operator defined as

$$Q^{x=0} = [d^\dagger \cdot \tilde{s} + s^\dagger \cdot \tilde{d}]^{(2)} \tag{17}$$

The Hamiltonian corresponding to the cubic quadrupole operator with coupling parameter K is given by

$$H_Q = -K[\hat{Q}\hat{Q}\hat{Q}]^{(0)} \tag{18}$$

The expectation value of the Hamiltonian $H_Q$ is obtained by using the intrinsic coherent state (8) to yields

$$E_Q(N,\beta,\gamma) = -K\sqrt{\frac{8}{35}} \left[\frac{3N(N-1)}{(1+\beta^2)^2} + \frac{4N(N-1)(N-2)}{(1+\beta^2)^3}\beta^3 \cos 3\gamma\right] \tag{19}$$

The rigid triaxial is occur at γ=30o, which predicts a relations between the first three excited state energies as

$$\Delta E = E_{3_\gamma^+} - E_{2_\gamma^+} - E_{2_g^+} = 0 \tag{20}$$

$$E(3) = E(2_1^+) + E(2_2^+) \tag{21}$$

To remove the dependence on the boson number N in classical limit, we rewrite the Hamiltonian $H_Q$ in the form

$$H_Q = -\frac{K}{N^3}[QQQ]^{(0)} \tag{22}$$

The corresponding PES reads

$$E_Q(N,\beta,\gamma) = -4K\sqrt{\frac{8}{35}}\left[\frac{\beta^3 \cos 3\gamma}{(1+\beta^2)^3}\right] \tag{23}$$

Adding the eigenvalue $E_Q$ to the eigenvalue $E(\beta)$ equation (14) for the O(6) dynamical symmetry, yield

$$E(\beta, \gamma) = \frac{1}{(1+\beta^2)^3}[(\lambda - a_0)\beta^2 + (2\lambda - a_0)\beta^4 - 4K\sqrt{\frac{8}{35}}\ \beta^3 \cos 3\gamma + \lambda\beta^6] + \frac{1}{4}a_0 \quad (24)$$

It is clear that the resultant PES depends on $\gamma$ and leads to triaxiality.

For $\lambda=1$, $a_0=0$, $k>0$ or $k<0$ and $\gamma=0$ or $60°$, the PES exhibit a spherical and deformed shapes as illustrated in Figure (3i), while for $a_0<\lambda$, a spherical shapes occur as seen in Figure (3ii) and for $a_0>\lambda$, a deformed shape occur as in Figure (3iii). The critical point occur at $a_0= \lambda =1$ as in Figure (3iv).

Besides the excitation energies and the PES's, the B(E2) transition probabilities can be calculated using the O(6) IBM electric quadrupole operator $Q^{x=0}$,

$$T(E2) = \alpha Q^{x=0} \quad (25)$$

The reduced E2 transition probabilities reads

$$B(E2, I_i \rightarrow I_f) = \frac{|\langle I_f|T(E2)|I_i\rangle|^2}{2I_i+1} \quad (26)$$

## 4. Outline of the Collective Nuclear Softness Model (CNSM)

The energy expression formula for axially symmetric rigid rotor is given by

$$E(I) = \frac{\hbar^2}{2J}I(I + 1) \quad (27)$$

This however, always predicts states lying higher than that given by experiments. If we attribute this effect to the variation of the moment of inertia J with angular momentum I, then we can write equation (27) after introducing the variable moment of inertia $J_I$ as

$$E(I) = \frac{\hbar^2}{2J_I}I(I + 1) \quad (28)$$

If we make the Taylor series expansion of the variable moment of inertia $J_I$ about the ground state Jo for I=0, then yield

$$J_I = J_0(1 + \sigma_1 I + \sigma_2 I^2 + \sigma_3 I^3 + \cdots.) \quad (29)$$

where the nuclear softness σ is an appropriate parameter to study the variation in moment of inertia. It is defined as the relative increase of the moment of inertia with angular momentum.

$$\sigma_n = \frac{1}{n! J_I} \left(\frac{\partial^n J_I}{\partial I^n}\right)_{I=0} \qquad (30)$$

Keeping the nuclear softness to only first order, that is putting σ$_2$ , σ$_3$ ,…equal to zero, the equation of energy reduced to a two parameters expression

$$E(I) = \frac{\hbar^2}{2J_0} \frac{I(I+1)}{(1+\sigma I)} \qquad (31)$$

This expression is denoting as the two parameters collective nuclear softness model (CNS2)[41,42] where J$_o$ and σ are the constant parameters.

## 5. Odd-Even Spin Energy Staggering in γ-Band

Energy staggering patterns between odd and even spin sequences were investigated and interpreted in several literatures [53-56]. In this paper, to exhibit this staggering phenomenon in γ-bands, we used the CNS2 model and suggested three different dimensionless staggering indices. The first two staggering indices S(I) and ΔE(I) depends on the dipole transitions linking the two families of spins, while the third one Y(I) depends on the dipole transitions between the two families and the quadrupole transitions within each spin family, namely

$$S(I) = 1 - \frac{(I+1)E(I-1) + IE(I+1)}{(2I+1)E(I)} \qquad (32)$$

$$\Delta E(I) = \frac{3}{E(2_g^+)} [E(I) - 2E(I-1) + E(I-2)] \qquad (33)$$

$$Y(I) = \left(\frac{4I-2}{2I}\right) \frac{E(I) - E(I-1)}{E(I) - E(I-2)} - 1 \qquad (34)$$

where E(I) is the energy of the state of angular momentum I.

Notice that, each point in the three staggering indices includes three consecutive energies. For axially symmetric pure rotator E(I)=AI(I+1), the two indices S(I) and Y(I) equal zero while the ΔE(I) index is equal to one. We say that odd-even staggering is observed if the staggering index exhibits a zigzag curve with increasing spin. The S(3), S(5) and ΔE(4) are indices of special interest

$$S(3) = 1 - \frac{4E(2) + 3E(4)}{7E(3)} \qquad (35)$$

$$S(5) = 1 - \frac{6E(4) + 5E(6)}{11E(5)} \tag{36}$$

$$\Delta E(4) = \frac{3}{E(2_g^+)}[E(4) - 2E(3) + E(2)] \tag{37}$$

## 6. Numerical Calculations and Discussion Applied to Xenon Isotopic Chain

The Xe isotopes have proton number Z=54 with 4 protons above the magic number 50 given 2 proton boson number, while the neutron number in the $^{120-134}_{54}Xe$ isotopic chain are between N=66 an N=80, so that the doubly closed shell of Z=50 and N=82 is assumed such that the valence neutrons are treated as holes, whereas the protons are valence particles, the corresponding number of bosons starts from $N_B$=10 in $^{120}$Xe to $N_B$=3 in $^{134}$Xe.

The important quantities which can characterize the collective structures and the shape phase transitions are the low-lying energy ratios and the electric quadrupole transition rates denoted as; $R_{4_1^+/2_1^+} = E(4_1^+)/E(2_1^+)$, $R_{6_1^+/2_1^+} = E(6_1^+)/E(2_1^+)$, $R_{2_2^+/2_1^+} = E(2_2^+)/E(2_1^+)$, $R_{3_2^+/2_1^+} = E(3_1^+)/E(2_1^+)$, $R_{4_2^+/2_1^+} = E(4_2^+)/E(2_1^+)$, $R_{0_2^+/2_1^+} = E(0_2^+)/E(2_1^+)$ and $B_{I_f/I_i} = \frac{B(E2;I_f \to I_i)}{B(E2;2_1^+ \to 0_1^+)}$

The IBM calculations have been done by using the code PHINT [57] and a simulated search program and the root mean square (rms) fits of energies have been performed using the standard $\chi$ measure

$$\chi = \sqrt{\frac{1}{N}\sum_{i=1}^{N}(\frac{E^{exp}(i) - E^{cal}(i)}{E^{exp}(i)})^2}$$

where N is the number of energy levels fitted for the Xe isotopes. The same procedure for B(E2) transition rates have been used.

The values of energy ratios $R_{I/2}$ and the electric quadrupole transition rates $B_{If/Ii}$ are calculated for our $^{120-134}$Xe isotopic chain and compared to those of the available experimental data [58] and the results are listed in Tables [1.2]. The ratio $R_{4_2^+/2_1^+}$ is the best signature to exhibit the shape transition; it has a limiting value 2 for quadrupole vibrator and the value 2.5 for a non-axial γ-soft rotator. As it seen from Table(1) ratio $R_{4_2^+/2_1^+}$ increase for about 2.1 for $^{134}$Xe to about 2.5 for $^{120}$Xe. Our calculated energy levels reflect the triaxialily and softness in Xe isotopes because they verify the conditions $\Delta E_1 = E(3_1^+) - [E(2_1^+) + E(2_2^+)]$ for triaxial nucleus,

$\Delta E_2 = E(3_1^+) - [2E(2_1^+) + E(4_1^+)]$ for γ-soft nucleus. The difference $\Delta E_1$ is low for $^{126,128}$Xe while the difference $\Delta E_2$ is large which reflects the triaxial nature. In $^{122}$Xe, $\Delta E_1$ is large and $\Delta E_2$ is small which show γ-soft nature.

Figure (4) illustrate the PES's as a function of deformation parameter β for the $^{120-134}$Xe isotopic chain calculated from equation (14) before adding the three body interaction. The parameters $a_0$ and λ are fixed at $a_0$=101.2KeV and λ=156.2KeV for all isotopes, while the values of the parameter $\epsilon$ are listed in Table (3). We see that the nuclei in Xe isotopic chain evolve from spherical vibrator U(5) to γ-soft rotor O(6) by increasing the boson number from N=3 (heavy isotope $^{134}$Xe) to N=10 (light isotope $^{120}$Xe). The isotopes $^{126,128}$Xe are good candidates for second order critical nuclei as indicated in reference [59].

The PES's are calculated after adding the cubic term interaction $[QQQ]^{(0)}$ to introduce a degree of triaxiality. The result is illustrated in Figure (5). The calculated asymmetric (triaxiality) parameter γ is given at the upper part of each curve. We see that by increasing the value of the three body interaction, the spectrum depends on the boson number N, approach those of triaxiality at N=7 for isotope $^{126}$Xe beginning from the behavior isotope $^{134}$Xe (N=3).

The energy spectra of γ-soft bands in $^{120-126}$Xe isotopes are determined by fitted the experimental energy levels to the calculated ones extracted from the suggested CNS2 model by using a second simulated search program to minimize the rms deviation

$$\text{Dev} = \sqrt{\frac{1}{N}\sum_{i=1}^{N}\left(\frac{E_{\gamma b}^{Exp}(I_i) - E_{\gamma b}^{CNS}(I_i)}{E_{\gamma b}^{Exp}(I_i)}\right)^2}$$

where N is the number of observed energies entering the fitting procedure. The optimized best fit parameters $J_0$ and σ corresponding to odd and even spin families for each nucleus are listed in Table(4). A comparison between the experimental spectra of γ-bands in $^{120-126}$Xe and the corresponding CNS2 model are illustrated in Figure (6), the agreement between them is excellent. Using the values of energy levels of γ–bands calculated in framework of CNS2 model, the three staggering indices S(I),ΔE(I) and Y(I) between odd-spin family are calculated and illustrated in Figure(7) as functions of spin. A clear staggering pattern is observed.

**Conclusion**

In the present paper, we have presented two models: the modified interacting boson model (IBM) and the collective nuclear softness (CNS) model. The IBM to study the shape phase transition U(5)-triaxiality-O(6) in ground state bands of $^{120-134}$Xe isotopic chain, while the CNS model to exhibit the ΔI=1 odd-even spin energy staggering in γ-bands of $^{120-126}$Xe isotopes. We added to the O(6) IBM Hamiltonian the effective three-body interaction $[\hat{Q}\hat{Q}\hat{Q}]^{(0)}$ where $\hat{Q}$ is the O(6) IBM quadrupole operator. By using a computer simulated search programs and the code PHINT-IBM a fitting procedure for each nucleus is performed to get the best parameters of our suggested models, in order to obtain a minimum root mean square deviation between the experimental and calculated energies and B(E2) transition rates for some selected low-lying levels. The potential energy surfaces (PES,s) are calculated and analyzed by using the method of intrinsic coherent states and the critical points are identified. The γ-soft rotation phase has been arised in light Xe isotopes like the nucleus $^{122}$Xe with energy ratio $R_{4/2}$ near 2.5, while the spherical vibrator phase has been arise in heavy isotopes like the nucleus $^{134}$Xe with $R_{4/2}$=2.04. A degree of triaxiality has been observed in $^{126,128}$Xe isotopes. In framework of CNS model, three staggering indices have been suggested. We proved that the even spin numbers I=2,4,6,8,10 of γ-bands are depressed with respect to the odd-spin ones I=3,5,7,9 for the four isotopes $^{120-126}$Xe.

## Acknowledgment

The authors extend their appreciation to the Deputyship for Research & Innovation, Ministry of Education in Saudi Arabia for funding this research work through the project number (IF2/PSAU/2022/ 01/23196)

Table (1) Comparison of the IBM calculations with those of the experimental ones for the energy ratios $R_{I_i/2_1}$ of low-lying levels in the $^{120-134}$Xe isotopic chain

| Nuclide | | $R_{4_1^+/2_1^+}$ | $R_{6_1^+/2_1^+}$ | $R_{2_2^+/2_1^+}$ | $R_{3_1^+/2_1^+}$ | $R_{4_2^+/2_1^+}$ | $R_{0_2^+/2_1^+}$ |
|---|---|---|---|---|---|---|---|
| $^{120}$Xe | Exp | 2.4678 | 4.3312 | 2.7156 | 3.9418 | 4.3437 | 2.8135 |
| | Cal | 2.5199 | 4.0699 | 2.7920 | 4.0546 | 4.2711 | 3.0253 |
| $^{122}$Xe | Exp | 2.5009 | 4.4283 | 2.5450 | 3.6655 | 4.2341 | 3.4688 |
| | Cal | 2.5647 | 4.1774 | 2.5793 | 3.9501 | 4.2006 | 3.4644 |
| $^{124}$Xe | Exp | 2.4826 | 4.3737 | 2.3910 | 3.5240 | 4.0616 | 3.5841 |
| | Cal | 2.5564 | 4.2602 | 2.4212 | 3.7931 | 4.0135 | 3.4269 |
| $^{126}$Xe | Exp | 2.4238 | 4.2070 | 2.2640 | 3.3905 | 3.8298 | 3.3807 |
| | Cal | 2.4913 | 4.1803 | 2.2994 | 3.5667 | 3.8761 | 3.2569 |
| $^{128}$Xe | Exp | 2.3326 | 3.9224 | 2.1888 | 3.2276 | 3.6203 | 3.5740 |
| | Cal | 2.3799 | 4.1090 | 2.2470 | 3.3505 | 3.5776 | 3.2941 |
| $^{130}$Xe | Exp | 2.2471 | 3.6266 | 2.0932 | 3.0454 | 3.3730 | 3.3456 |
| | Cal | 2.3379 | 3.8207 | 2.0449 | 3.0619 | 3.3649 | 3.2979 |
| $^{132}$Xe | Exp | 2.1570 | 3.1628 | 1.9438 | 2.1294 | 2.3174 | |
| | Cal | 2.2369 | 3.2999 | 1.9109 | 2.1701 | 2.2955 | |
| $^{134}$Xe | Exp | 2.0437 | 2.5224 | 1.9051 | | | |
| | Cal | 2.1399 | 2.7999 | 1.8801 | | | |

Table (2) Comparison of the IBM calculations with those of the available experimental ones for the electric quadrupole transition rates $B_{I_i/2_1}$. The prediction of the O(6) limit is also given.

| Nuclide | | $B_{4/2}$ | $B_{6/4}$ | $B_{8/6}$ | $B_{10/8}$ |
|---|---|---|---|---|---|
| $^{124}$Xe $N_B=8$ | Cal | 1.501 | 1.991 | 2.490 | 3.354 |
| | Exp | 1.34(24) | 1.59(71) | 0.63(29) | 0.29(8) |
| | O(6) | 1.354 | 1.458 | 1.420 | 1.282 |
| $^{124}$Xe $N_B=6$ | Cal | 1.790 | 2.853 | 4.232 | 6.001 |
| | Exp | 1.47(2) | 1.94(26) | 2.39(4) | 2.74(114) |
| | O(6) | 1.309 | 1.333 | 1.181 | 0.897 |
| $^{124}$Xe $N_B=4$ | Cal | 2.990 | 2.032 | 17.492 | |
| | Exp | 1.24(18) | | | |
| | O(6) | 1.205 | 1.666 | 0.625 | |

Table (3) The values of the parameter $\varepsilon$ ($a_0$=101.2 Kev, $\lambda$=156.2 KeV) for the $^{134-120}$Xe isotopic chain

| | $^{134}$Xe | $^{132}$Xe | $^{130}$Xe | $^{128}$Xe | $^{126}$Xe | $^{124}$Xe | $^{122}$Xe | $^{120}$Xe |
|---|---|---|---|---|---|---|---|---|
| N | 3 | 4 | 5 | 6 | 7 | 8 | 9 | 10 |
| E(KeV) | 727.4 | 674.2 | 621 | 567.8 | 514.6 | 461.4 | 408.2 | 355 |

Table (4) The adapted best parameters $J_o$ and $\sigma$ of the collective soft rotor model (CNS2) resulted from the fitting procedure for odd and even spin values for γ-bands in $^{120-126}$Xe nuclei.

| Nucleus | Sequence(I) | $J_o$ ($\hbar^2$ MeV$^{-1}$) | $\sigma$ |
|---|---|---|---|
| $^{120}$Xe | even | 0.45238 | 3.60855 |
| | odd | 0.18220 | 8.64394 |
| $^{122}$Xe | even | 1.35523 | 1.02893 |
| | odd | 0.94440 | 1.53418 |
| $^{124}$Xe | even | 1.04045 | 1.35970 |
| | odd | 0.73187 | 1.95725 |
| $^{126}$Xe | even | 0.88975 | 1.56287 |
| | odd | 0.73592 | 1.85669 |

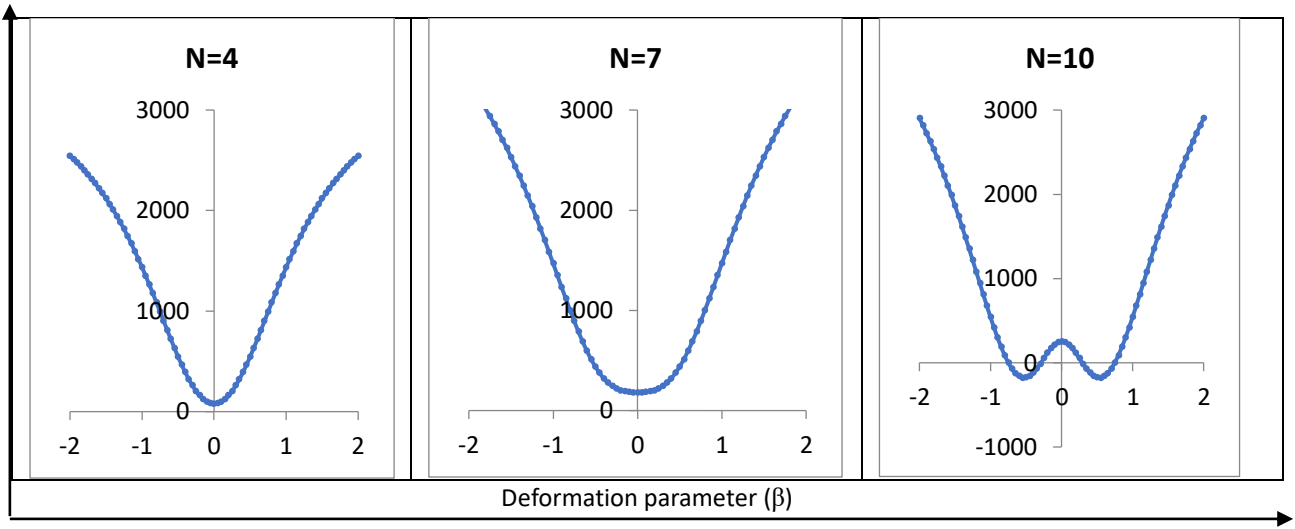

Figure (1) The calculated PES′s as a function of deformation parameter β for O(6) IBM Hamiltonian equation (1) at $a_0$=101.2 and λ varying for produce critical point at N=7.

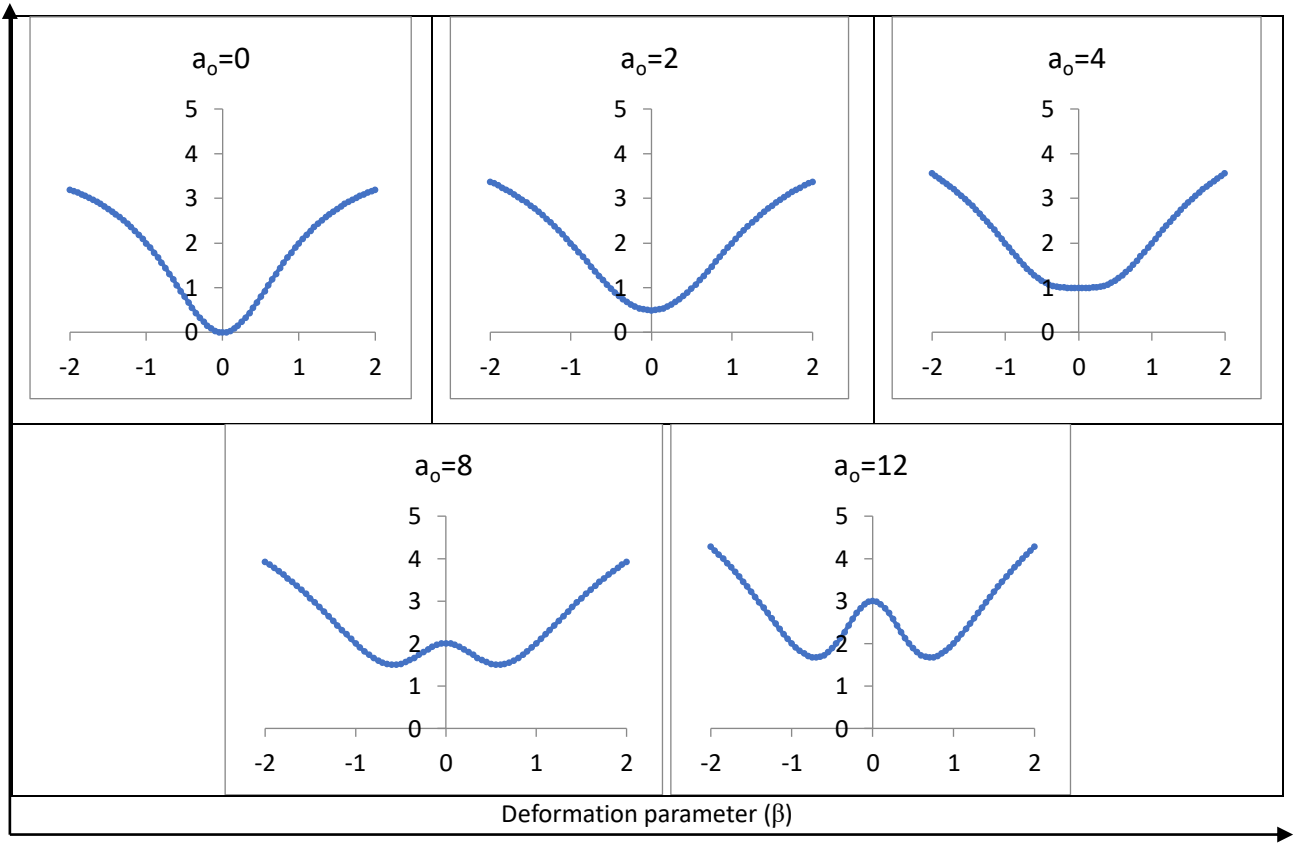

Figure (2) The calculated PES′s versus deformation parameter β in classical limit corresponding to Hamiltonian ( 1 ) for λ =4 and a₀=0,2,4,8,12.

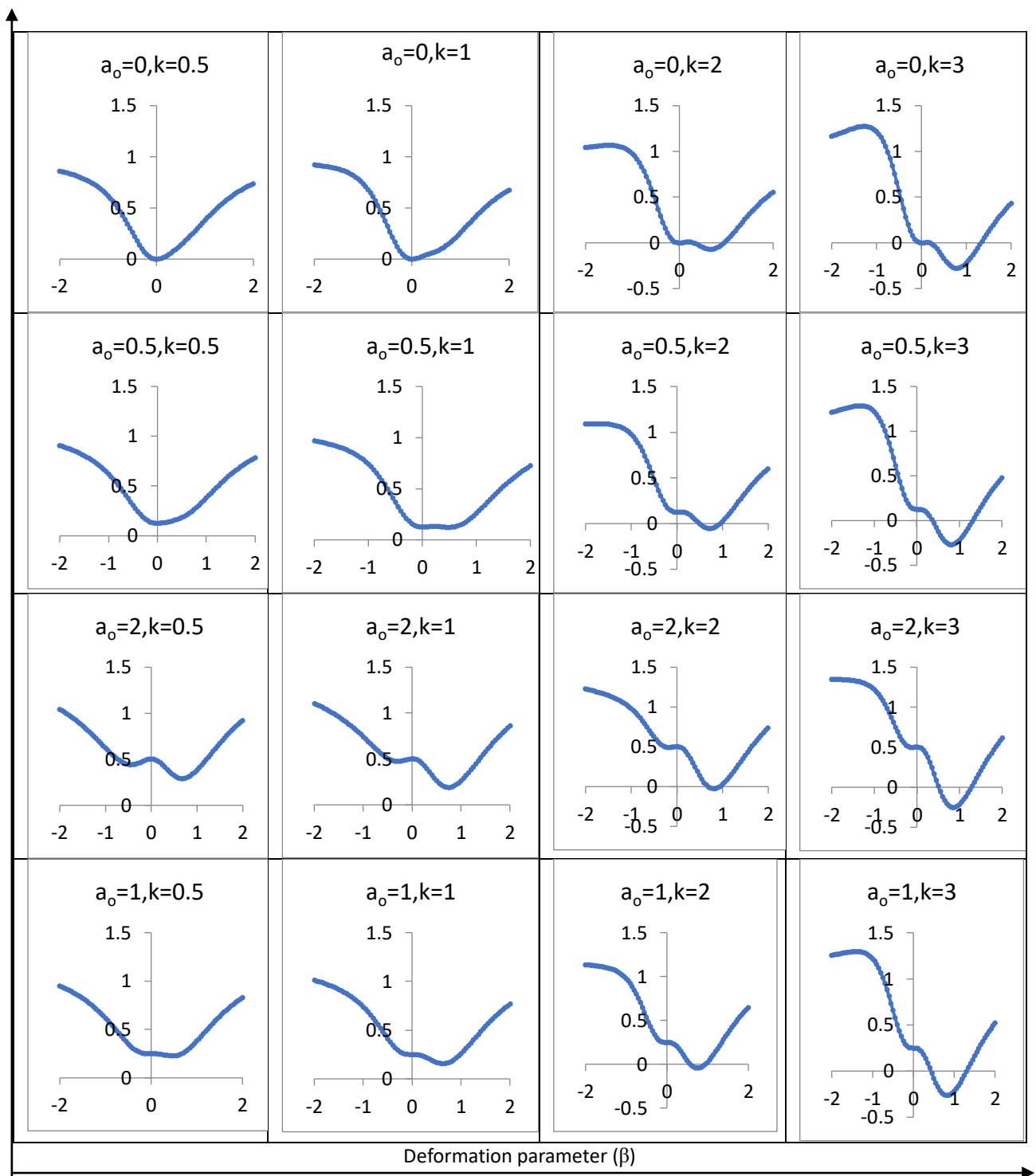

Deformation parameter (β)

Figure (3) The calculated PES′s equation versus deformation parameter β for ө=0º , ө =60º and λ=1 for different values of the parameters **a** and **k**.

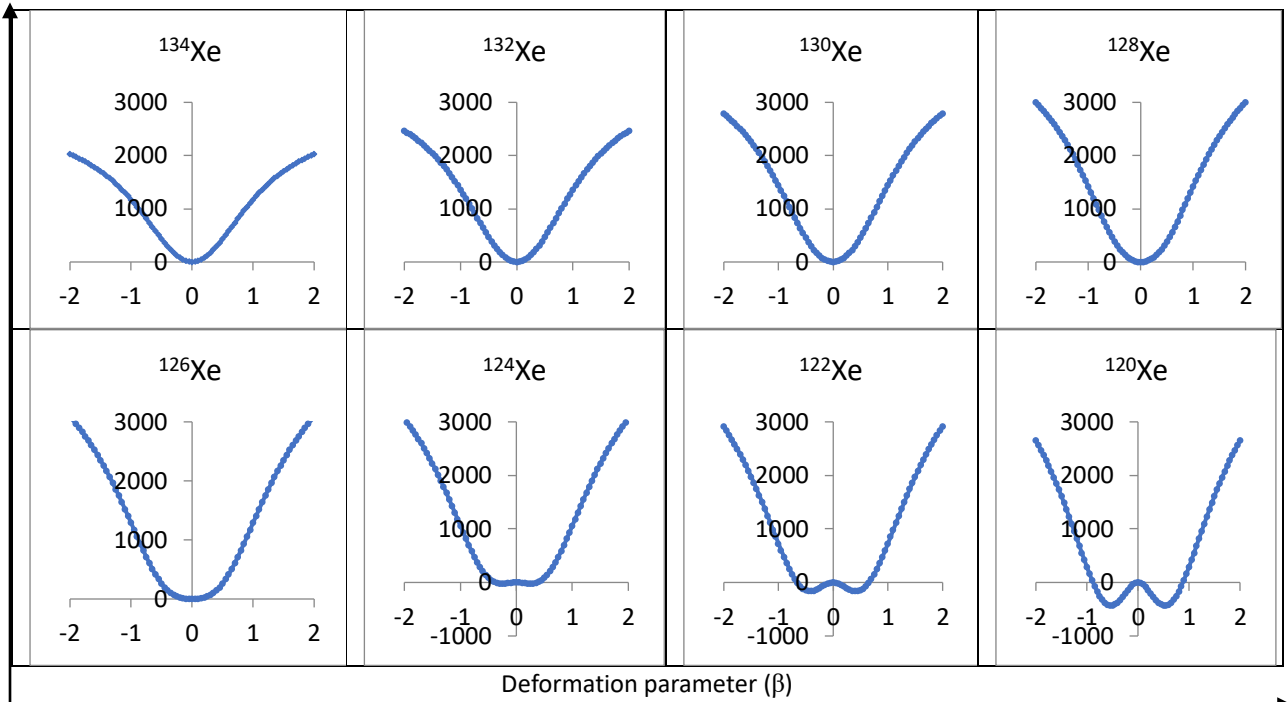

Figure (4) The calculated PES's using IBM versus deformation parameter β before adding the cubic term interaction. The critical points in the transition from U(5) (spherical case) to O(6) ( γ-unstable deformed case) are at N=6,7 for $^{128}$Xe , $^{126}$Xe

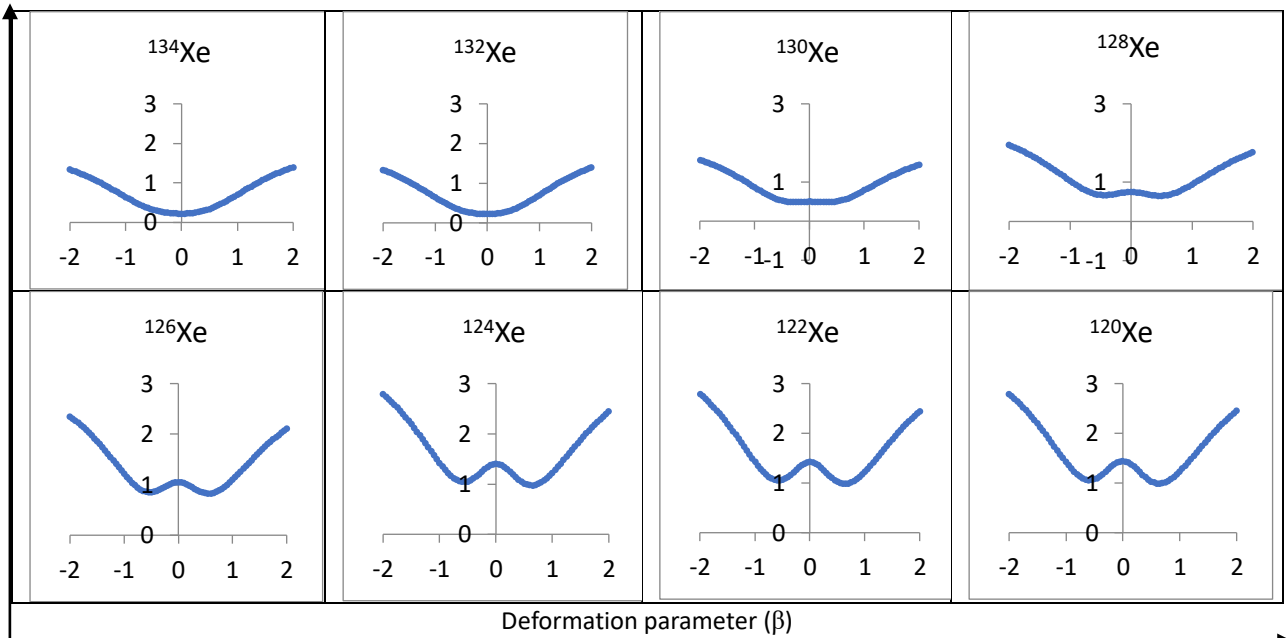

Figure (5) The calculated PES,s using IBM versus deformation parameter β after adding the cubic term interaction

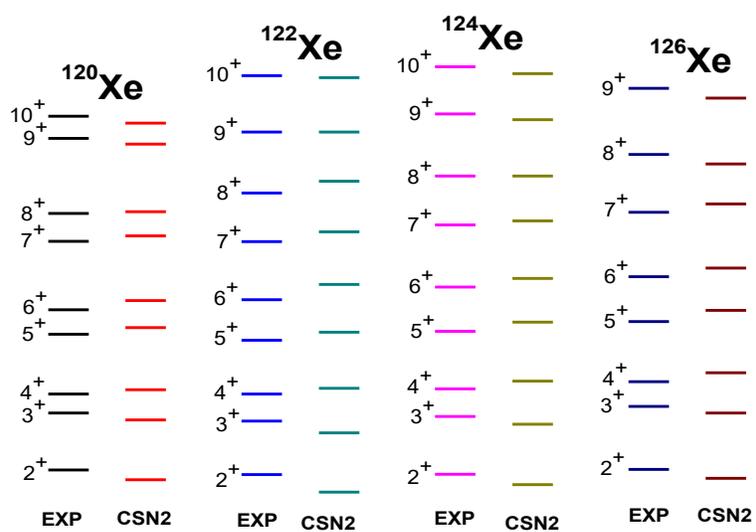

Figure (6) Comparison of experimental energy levels with the calculated ones (in framework of CNS2 model) for γ-bands in $^{120-126}$Xe isotopes

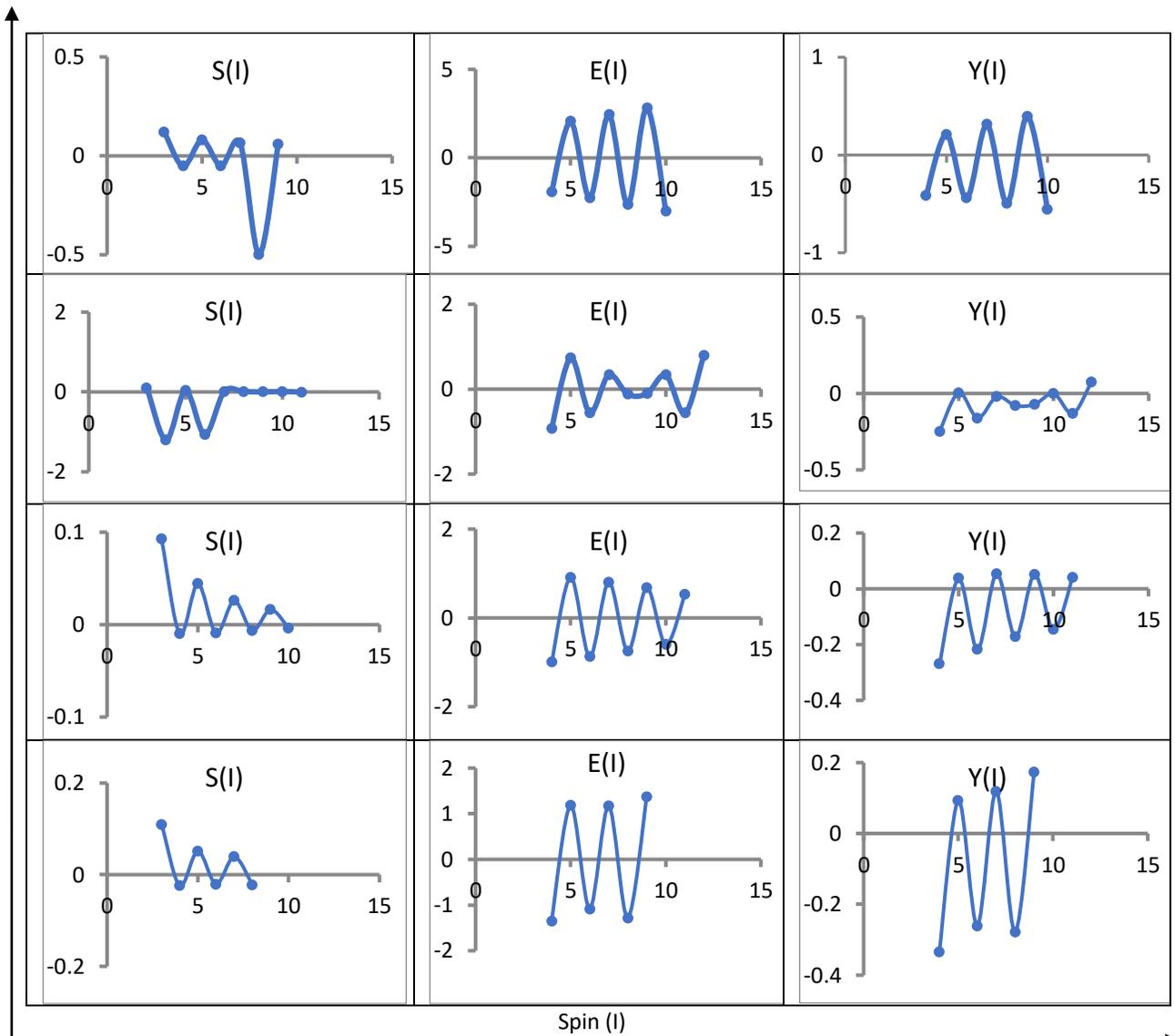

Figure (7) The CNS2 model calculations of ΔI=1 odd-even spin energy staggering indices S(I), ΔE(I) and Y(I) versus nuclear spin I for $^{120-126}$Xe isotopes.